\documentclass[prl,twocolumn,superscriptaddress,showpacs]{revtex4-1}

\usepackage{graphicx}
\usepackage{color}
\usepackage{amsmath}
\usepackage{amsfonts}
\usepackage{amssymb}
\usepackage{bm}
\usepackage{soul}
\usepackage{hyperref}

\begin{document}
%%%%%%%%%%%%%%%%%%%%%%%%%%%%%%%%%%%%%%%%%%%%%%%%%%%%%%%%%%%%

\title{Chiral Bogoliubons in Nonlinear Bosonic Systems}

\author{Charles-Edouard Bardyn}
\affiliation{Institute for Quantum Information and Matter, Caltech, Pasadena, California 91125, USA}
\author{Torsten Karzig}
\affiliation{Institute for Quantum Information and Matter, Caltech, Pasadena, California 91125, USA}
\author{Gil Refael}
\affiliation{Institute for Quantum Information and Matter, Caltech, Pasadena, California 91125, USA}
\author{Timothy C. H. Liew}
\affiliation{Division of Physics and Applied Physics, Nanyang Technological University 637371, Singapore}

\begin{abstract}
We present a versatile scheme for creating topological Bogoliubov excitations in weakly interacting bosonic systems. Our proposal relies on a background stationary field that consists of a Kagome vortex lattice, which breaks time-reversal symmetry and induces a periodic potential for Bogoliubov excitations. In analogy to the Haldane model, no external magnetic field or net flux is required. We construct a generic model based on the two-dimensional (2D) nonlinear Schr\"odinger equation and demonstrate the emergence of topological gaps crossed by chiral Bogoliubov edge modes. Our scheme can be realized in a wide variety of physical systems ranging from nonlinear optical systems to exciton-polariton condensates.
\end{abstract}


%03.75.Lm 	Tunneling, Josephson effect, Bose-Einstein condensates in periodic potentials, solitons, vortices, and topological excitations (see also 74.50.+r Tunneling phenomena; Josephson %effects in superconductivity)
%42.65.-k 	Nonlinear optics
%71.36.+c 	Polaritons (including photon-phonon and photon-magnon interactions)
%67.85.Hj 	Bose-Einstein condensates in optical potentials

%71.35.-y		Excitons and related phenomena
%85.75.-d 	Magnetoelectronics; spintronics: devices exploiting spin polarized transport or integrated magnetic fields
%42.70.Qs 	Photonic bandgap materials (for photonic crystal lasers, see 42.55.Tv)
%03.65.Vf 	Phases: geometric; dynamic or topologica
%73.63.-b 	Electronic transport in nanoscale materials and structures (see also 73.23.-b Electronic transport in mesoscopic systems
%03.75.Mn 	Multicomponent condensates; spinor condensate

\maketitle

%%%%%%%%%%%%%%%%%%%%%%%%%%%%%%%%%%%%%%%%%%%%%%%%%%%%%%%%%%%%

%========================================================================================
% Introduction
%========================================================================================

{\it Introduction.---}The quantum Hall effect is one of the most celebrated results of modern condensed matter physics~\cite{Klitzing1980}. The robustness of the Hall conductance can be traced back to the non-trivial topology of the underlying electronic band structure~\cite{Thouless1982}, which ensures the existence of chiral edge modes and thus eliminates backscattering. Recently there was a surge of interest in the possibility to exploit such topology to create chiral bosonic modes in driven-dissipative systems --- with possible applications for one-way transport of photons~\cite{Lu2014,Haldane2008,Wang2009,Koch2010,Hafezi2011,Poo2011,Umucalilar2011,Fang2012,Hafezi2013a,Jia2013}, polaritons~\cite{Karzig2014,Nalitov2014,Bardyn2014}, excitons~\cite{Yuen-zhou2014,Bardyn2014}, magnons~\cite{Shindou2013,Zhang2013}, and phonons~\cite{Peano2014,Yang2014}. A common thread through these seemingly diverse ideas has been to induce topology by external manipulations of a single-particle band structure, with interactions playing a negligible role. Exceptions from this non-interacting paradigm are proposals that combine strong interactions with externally induced artificial gauge fields to create non-equilibrium analogs of bosonic fractional quantum Hall states~\cite{Cho2008,Umucalilar2012,Hayward2012,Hafezi2013}.

Here we take a new perspective and consider (bosonic) Bogoliubov excitations (``Bogoliubons'') where \emph{weak interactions} induce a non-trivial topology~\footnote{Topological Bogoliubov excitations have recently been proposed in a one-dimensional system~\cite{Engelhardt2015}. Although interactions can change the topology, in that case the latter results from spin-orbit coupling combined with the inversion symmetry of the problem.}. We demonstrate that topological Bogoliubons naturally arise on top of a condensate that exhibits a lattice of vortex-antivortex pairs, with no net flux required. Interactions are key to harness the time-reversal symmetry breaking provided by the condensate vortices. From the point of view of Bogoliubov excitations, they generate non-trivial ``hopping'' phases which lead to an analog of the Haldane lattice model~\cite{Haldane1988}. The periodic potential defining the corresponding lattice can be introduced as an external potential or generated by interactions with the underlying condensate.

%%%%%%%%%FIGURE%%%%%%%%%
\begin{figure}[th]
    \begin{center}
        \includegraphics[width=0.96\linewidth]{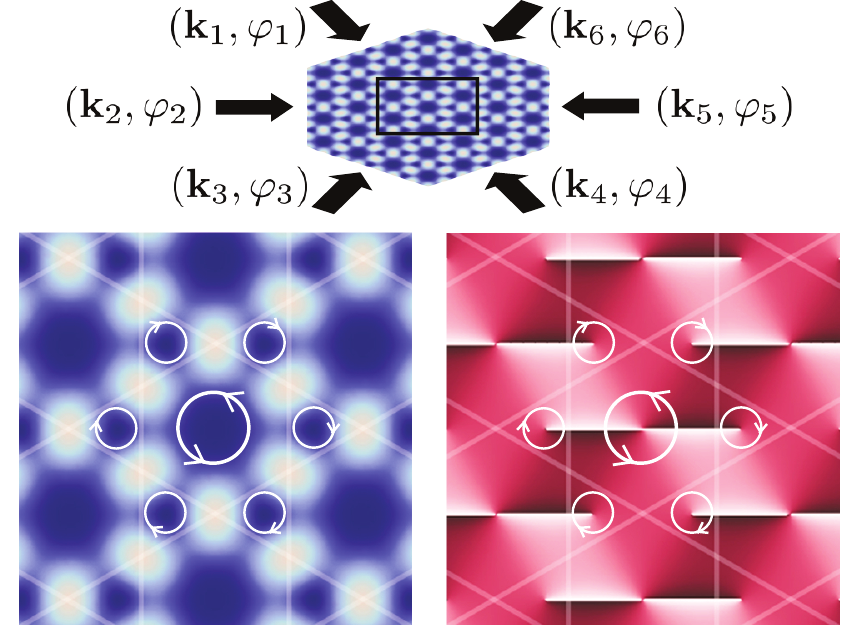}
        \caption{Top: Optical pumping scheme allowing to create a suitable condensate for topological Bogoliubons, with incident field composed of six equal-frequency plane-wave components with wave vectors $\mathbf{k}_n$ as depicted by arrows and phases chosen as $\varphi_n = 0$ except for $\varphi_4 = 2\pi/3$ and $\varphi_6 = -2\pi/3$. Bottom: Intensity (left) and phase (right) pattern of the resulting field. Intensity maxima form a Kagome structure (white lines) with vortex-antivortex pairs located such that each hexagonal plaquette is threaded by a flux $2\Phi_0 = 4\pi$, with smaller triangular plaquettes threaded by $-\Phi_0$. Although fluxes cancel out over the whole system, a well-defined chirality emerges, defined by the sign of $\Phi_0$. Bogoliubov excitations of the condensate experience non-trivial fluxes due to interactions, leading to TR symmetry breaking and ultimately to topological states.}
        \label{fig:vortexLattice}
    \end{center}
\end{figure}
%%%%%%%%%%%%%%%%%%%%%%

Although our scheme can be applied to any system described by a 2D nonlinear Schr\"odinger equation (Gross-Pitaevskii equation, or analog thereof), we focus on systems of weakly interacting bosons that have a light component, where the required vortex lattice can readily be obtained from the interference of several coherent (optical) fields (see Fig.~\ref{fig:vortexLattice}). In this setting, the phase-imprinting mechanism through which non-trivial topology is ultimately granted is analogous to that proposed a few years ago in the context of optomechanical systems~\cite{Habraken2012}. The same mechanism was recently applied to create topological phonons using photons that remain trivial~\cite{Peano2014}. Here we demonstrate that vortex lattices can be exploited in a much broader variety of systems where, in contrast to previous works, topology does not emerge from the coupling to a different bosonic species, but is intrinsically enabled by interactions.

Our analysis starts with the 2D nonlinear Schr\"odinger equation commonly used to describe weakly nonlinear dispersive physical systems:
\begin{equation}
    i \partial_t \psi(\mathbf{x},t) = \left( \omega_d(-i \boldsymbol{\nabla}) + \alpha |\psi(\mathbf{x},t)|^2 + V(\mathbf{x}) \right) \psi(\mathbf{x},t),
    \label{eq:NLSE}
\end{equation}
where $\psi(\mathbf{x},t)$ is a complex field (or ``wavefunction'') describing the amplitude and phase of a coherent field, and $\omega_d(-i \boldsymbol{\nabla})$ describes the dispersion of the system. Nonlinearities are characterized by a parameter $\alpha$ which may be positive (repulsive/self-defocusing interactions) or negative (attractive/self-focusing interactions). We also allow for a spatially dependent potential $V(\mathbf{x})$ although, as we will show, this is only essential when $\alpha > 0$. Equation~\eqref{eq:NLSE} is applicable to a wide variety of physical systems ranging from light propagation through Kerr nonlinear media~\cite{Kelley1965} to the dynamics of exciton-polariton systems~\cite{Carusotto2013} and Bose-Einstein condensates~\cite{Leggett2001}. Below we outline a generic mechanism for creating topological Bogoliubov excitations, assuming that the underlying bosonic fields have a light component. We then discuss specific implementations.

%========================================================================================
% Theoretical scheme
%========================================================================================

{\it Theoretical scheme.---}The first and essential ingredient of our scheme consists of a stationary field $\psi_0(\mathbf{x},t)$ which exhibits vortex-antivortex pairs and intensity maxima that form a Kagome lattice pattern, as illustrated in Fig.~\ref{fig:vortexLattice}. In bosonic systems with a photonic component, such a Kagome ``vortex lattice'' can be directly imprinted onto the field $\psi(\mathbf{x},t)$ using an optical coherent field (or ``pump'') composed of six plane-wave components as depicted in Fig.~\ref{fig:vortexLattice} (see also Supplementary Material~\cite{supplementalMaterial}). Denoting by $\hat{\Psi}_L(\mathbf{x})$ the (bosonic) field operator associated with the light component, such coherent pumping can be described by a Hamiltonian term of the form $\int \mathrm{d}\mathbf{x} f(\mathbf{x})e^{-i \omega_0 t} \hat{\Psi}^\dagger_L(\mathbf{x}) + h.c.$, where $f(\mathbf{x})$ and $\omega_0$ respectively denote the pump spatial profile and frequency. In the mean-field limit where Eq.~\eqref{eq:NLSE} applies, this results in an additional pumping term proportional to $f(\mathrm{x})e^{-i \omega_0 t}$ (see, e.g., Ref.~\cite{Ciuti2005}). The required pumping field is readily obtainable with a spatial light modulator~\cite{Tosi2012} or by passing light through a mask~\cite{Dall2014}.

Remarkably, the vortex-antivortex pairs arising in the above Kagome vortex lattice are located in such a way that each elementary hexagonal plaquette is threaded by a flux $2\Phi_0 = 4\pi$ (i.e., contains 2 vortices), while smaller triangular plaquettes are threaded by $-\Phi_0$ (i.e., contain a single antivortex) [see Fig.~\ref{fig:vortexLattice}]. Since the incident pumping field has no net orbital angular momentum to transfer to the internal field, vortices and antivortices always come in pairs and fluxes cancel out over the whole system. Nevertheless, the stationary field $\psi_0(\mathbf{x},t)$ breaks time-reversal symmetry, as revealed by the ``chirality'' defined by the sign of $\Phi_0$ (which is set by the phases of the pumping field~\cite{supplementalMaterial} and does not change under rotations or translations in the plane that leave the Kagome lattice of intensities invariant~\footnote{Such chirality would not appear in square or triangular lattices of vortex-antivortex pairs, since in that case fluxes would change sign upon rotation by $\pm \pi/2$ and $\pm 2\pi/3$, respectively.}).

The fact that $\psi_0(\mathbf{x},t)$ exhibits a staggered flux pattern with no net flux but a well-defined chirality is very reminiscent of Haldane's seminal proposal for quantum Hall physics with no external magnetic field~\cite{Haldane1988}. Here, however, the phase that an excitation would accumulate around any plaquette of the Kagome lattice (or around any loop in the plane) seems to be an integer multiple of $2\pi$ (equivalent to no phase at all), which would naively indicate that TR symmetry is preserved and that topological states are out of reach. Remarkably, we demonstrate below that Bogoliubov excitations, \emph{through interactions}, do experience the TR symmetry breaking of the pumping field. This will allow us to access a \emph{non-equilibrium interaction-induced} analog of Haldane's model.

To derive the spectrum of Bogoliubov excitations on top of $\psi_0(\mathbf{x},t)$, we perform a linearization of Eq.~\eqref{eq:NLSE}. We define the slowly-varying field $\phi(\mathbf{x},t) \equiv \psi(\mathbf{x},t) e^{i \omega_0 t}$, where $\omega_0$ is the pump frequency, and consider weak perturbations (or ``fluctuations'') of the form~\cite{Carusotto2004}
\begin{equation}
    \phi(\mathbf{x},t) = \phi_0(\mathbf{x}) + u(\mathbf{x}) e^{-i \omega t} + v^*(\mathbf{x}) e^{i \omega^* t},
    \label{eq:fluctuations}
\end{equation}
where $\phi_0(\mathbf{x})$ denotes the rotating-frame counterpart of the stationary field $\psi_0(\mathbf{x},t) \equiv \phi_0(\mathbf{x}) e^{-i \omega_0 t}$, $u(\mathbf{x})$ and $v(\mathbf{x})$ are (in general complex) functions determining the spatial form of the fluctuations, and $\omega$ is the frequency of the perturbations which is kept complex in order to capture potential instabilities~\cite{Carusotto2004}. Plugging the above expression into Eq.~\eqref{eq:NLSE} and neglecting second-order terms in $u(\mathbf{x})$ and $v(\mathbf{x})$ yields the Bogoliubov equation
\begin{align}
    \left(
    \begin{array}{cc}
        \omega'(\mathbf{x}) & \alpha \phi_0(\mathbf{x})^2 \\
        -\alpha \phi_0^*(\mathbf{x})^2 & -\omega'(\mathbf{x})
    \end{array}
    \right) \left(
    \begin{array}{c}
        u(\mathbf{x}) \\
        v(\mathbf{x})
    \end{array}
    \right) = \omega \left(
    \begin{array}{c}
        u(\mathbf{x}) \\
        v(\mathbf{x})
    \end{array}
    \right),
    \label{eq:bogoliubovEquation}
\end{align}
where $\omega'(\mathbf{x}) \equiv \omega_d(-i \boldsymbol{\nabla}) - \omega_0 + 2 \alpha |\phi_0(\mathbf{x})|^2 + V(\mathbf{x})$. This shows that Bogoliubov excitations experience both the intensity and phase pattern of the underlying field $\phi_0(\mathbf{x})$.

In the case of repulsive interactions ($\alpha > 0$), the effective potential term $2 \alpha |\phi_0(\mathbf{x})|^2$ tends to localize the excitations at the vortex/antivortex points of the lattice (see Fig.~\ref{fig:vortexLattice}). To allow for topological states, however, it is crucial that excitations remain localized away from vortices/antivortices so as to pick up a phase when hopping around them. This can be achieved, e.g., by introducing an external potential $V(\mathbf{x})$ with minima that coincide with the maxima of $|\phi_0(\mathbf{x})|^2$. In the case of attractive interactions ($\alpha < 0$), the effective potential $2 \alpha |\phi_0(\mathbf{x})|^2$ automatically localizes the excitations at the desired points, thus obviating the need for any additional potential.

According to Eqs.~\eqref{eq:fluctuations} and~\eqref{eq:bogoliubovEquation}, fluctuations described by $u(\mathbf{x})$ and $v(\mathbf{x})$ can be viewed as ``particle-hole'' analogs of each other: $u(\mathbf{x})$ as a ``particle'' excitation with dispersion $\omega'(\mathbf{x})$, and $v(\mathbf{x})$ as a ``hole'' excitation with opposite dispersion $-\omega'(\mathbf{x})$. The pump frequency $\omega_0$ sets the relative energy between the two. Interestingly, the minus sign present on the off-diagonal of Eq.~\eqref{eq:bogoliubovEquation} (a hallmark of bosonic Bogoliubov excitations~\cite{Leggett2001}) makes the corresponding matrix non-Hermitian. The stability of the system is then assessed by the imaginary part of $\omega$: when $\mathfrak{Im}(\omega) > 0$, fluctuations grow exponentially and $\phi_0(\mathbf{x})$ is an unstable solution of Eq.~\eqref{eq:NLSE}.

The off-diagonal coupling of Eq.~\eqref{eq:bogoliubovEquation} is crucial for our scheme: it harnesses the TR symmetry breaking encoded in the condensate phase and thus ultimately determines the size of the attainable topological gaps. The periodic potential $2 \alpha |\phi_0(\mathbf{x})|^2+V(\mathbf{x})$ also plays an important role: (i) it generates band crossings in the dispersion of $u$ (and similarly for $v$) at which gap openings are most easily obtained; and (ii) it ensures that Bogoliubov excitations ``live'' on the Kagome lattice pattern defined by the maxima of $|\phi_0(\mathbf{x})|^2$, which allows them to experience in an optimal way the fluxes induced by $\phi_0 (\mathbf{x})$.

%========================================================================================
% Tight-binding analysis
%========================================================================================

\begin{figure}[t]
    \begin{center}
        \includegraphics[width=\linewidth]{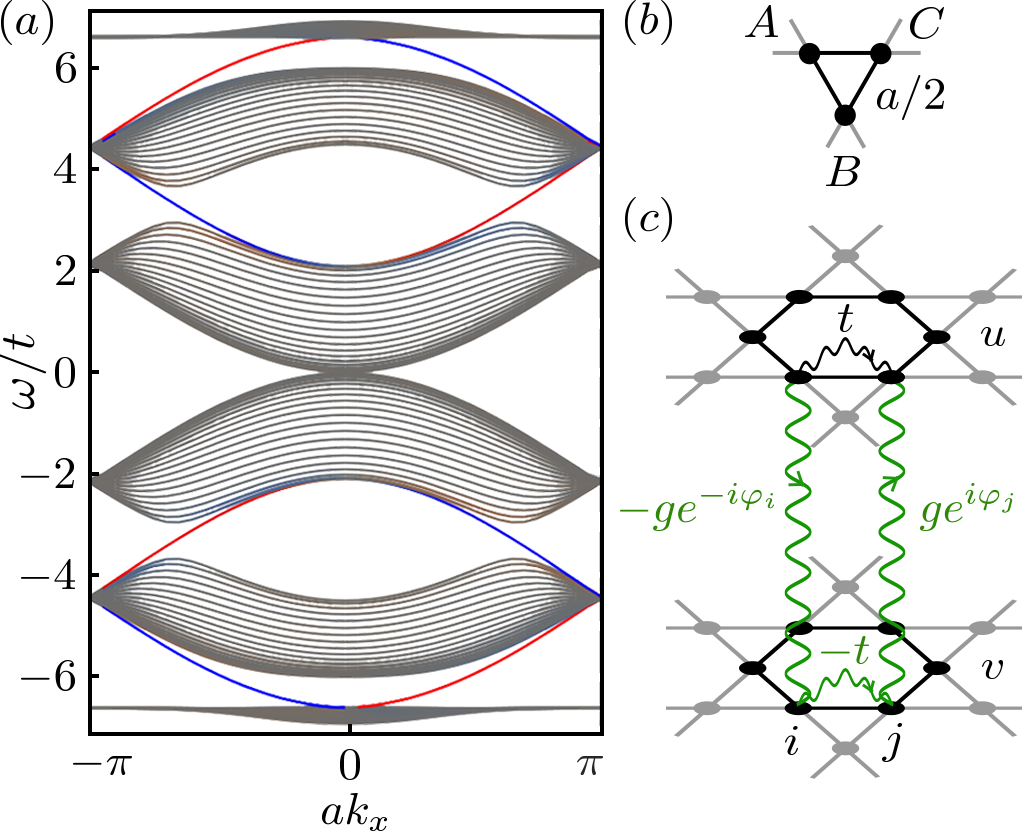}
        \caption{(a) Bogoliubov spectrum in the off-resonant tight-binding limit (in strip geometry): the particle-like (positive) part of the spectrum exhibits three bands with topological gaps crossed by pairs of counter-propagating chiral edge states, reminiscent of a Kagome lattice model with staggered fluxes~\cite{Ohgushi2000}. The red (blue) coloring indicates the degree of localization at the lower (upper) edge. Parameters were chosen as $\Omega/t = 6$ and $g/\Omega = 2/3$, in a strip geometry with periodic boundary conditions in the $x$-direction (see Supplemental Material~\cite{supplementalMaterial} for details). (b) Unit cell used in the tight-binding model, with phases for the on-site coupling chosen as $(0, 2\pi/3, -2\pi/3)$ in the basis $(A,B,C)$. (c) TR symmetry-breaking mechanism: off-resonant couplings between particle- and hole-like excitations ($u$ and $v$) lead to virtual transitions which renormalize the hopping term $t$.}
        \label{fig:tightBinding}
    \end{center}
\end{figure}

{\it Tight-binding analysis.---}Intuitive understanding of the Bogoliubov excitation spectrum can be gained from the tight-binding limit of Eq.~\eqref{eq:bogoliubovEquation}, which follows from the low-energy behavior obtained in a deep potential $2 \alpha |\phi_0(\mathbf{x})|^2 + V(\mathbf{x}) - \omega_0$ whose minima form a Kagome lattice. In the sector $u$ ($v$), we denote the corresponding on-site energy by $\Omega (-\Omega)$ and assume that excitations ``hop'' with amplitude $t(-t)$ between nearest-neighboring sites. The off-diagonal terms in Eq.~\eqref{eq:bogoliubovEquation} lead to couplings $\pm g e^{\pm i \varphi_j}$ between sectors $u$ and $v$ on each site $j$. To model the phase pattern of $\phi_0 (\mathbf{x})^2$, we choose the phases $\varphi_j$ so that each triangular unit cell of the Kagome lattice carries a single vortex (see Fig.~\ref{fig:tightBinding}b).

The mechanism giving rise to TR symmetry breaking (and, in turn, to topological states) can be revealed most clearly in the limit $2\Omega \gg g$ where particles and holes ($u$ and $v$) are coupled off-resonantly and weakly. In that case virtual transitions between particles and holes lead to a renormalization of the hopping amplitude and phase. In a perturbative picture~\footnote{Even though the matrix appearing in Eq.~\eqref{eq:bogoliubovEquation} is non-Hermitian, perturbation theory applies as usual, since non-Hermiticity only comes from the perturbative coupling between particles and holes.}, a particle-like Bogoliubov excitation can be turned into a hole-like excitation, hop to a neighboring site, and transform back into a particle-like excitation (see Fig.~\ref{fig:tightBinding}c). Since the particle-hole coupling involves a site-dependent phase, excitations pick up an overall phase $e^{\pm i (\varphi_j - \varphi_i)}$ (``$+$'' for particles and ``$-$'' for holes)~\footnote{This phase-imprinting mechanism is analogous to the one originally proposed in Ref.~\cite{Habraken2012} in optomechanical systems.}, which leads to an effective hopping
\begin{equation}
    t_\mathrm{eff} = \pm t \left( 1 + \left( \frac{g}{2\Omega} \right)^2 e^{\pm 2\pi i/3} \right)
    \label{eq:teff}
\end{equation}
in counter-clockwise direction around the triangular unit cell. The corresponding flux per triangular plaquette is given by $\Phi = \pm 3 \arctan(\sqrt{3}g^2/(8\Omega^2-g^2)) \neq 0$.

Equation~\eqref{eq:teff} illustrates one of the key aspects of our proposal: the fact that interactions generically lead to a non-trivial flux $\Phi$ (not a multiple of $\pi$), which breaks TR symmetry from the point of view of Bogoliubov excitations. In the off-resonant tight-binding regime considered here, particles and holes are both individually described by an effective Kagome lattice model with staggered flux pattern as depicted in Fig.~\ref{fig:vortexLattice}. If $\Phi$ was trivial (for $g = 0$), the positive or negative part of the Bogoliubov spectrum would exhibit three bands touching at high-symmetry points of the Brillouin zone (Dirac cones at K and K', and a quadratic band touching at $\Gamma$). The non-trivial flux $\Phi$ induced by interactions gaps out these degeneracies, leading to topological bands with a non-zero Chern number~\cite{Ohgushi2000} (see Fig.~\ref{fig:tightBinding}). To second order in $g$, the size of the topological gaps is given by~\cite{supplementalMaterial}
\begin{equation}
    \Delta_{K} = \frac{3tg^{2}}{4(\Omega-t)(\Omega+t/2)}, \ \Delta_\Gamma = \frac{3tg^{2}}{2(\Omega-t)(\Omega+2t)}.
\end{equation}
The Bogoliubov spectrum considered here in the off-resonant regime is stable, with only real eigenvalues.

%%%%%%%%%FIGURE%%%%%%%%%
%%%%%%%%%%%%%%%%%%%%%%

%========================================================================================
% Exciton-polariton systems
%========================================================================================

\emph{Exciton-polariton systems.---}An example of realization of our scheme can be practically considered with exciton-polariton condensates in semiconductor microcavities, which are renowned for their strong nonlinearities~\cite{Carusotto2013}. The Bogoliubov spectrum of polariton condensates has been measured in photoluminescence~\cite{Utsunomiya2008} and four-wave mixing~\cite{Kohnle2011} experiments. The required amplitude and phase pattern of $\phi_0(\mathbf{x})$ can be imprinted optically as described above~\cite{Liew2008}.

\begin{figure}[t]
    \begin{center}
        \includegraphics[width=0.96\linewidth]{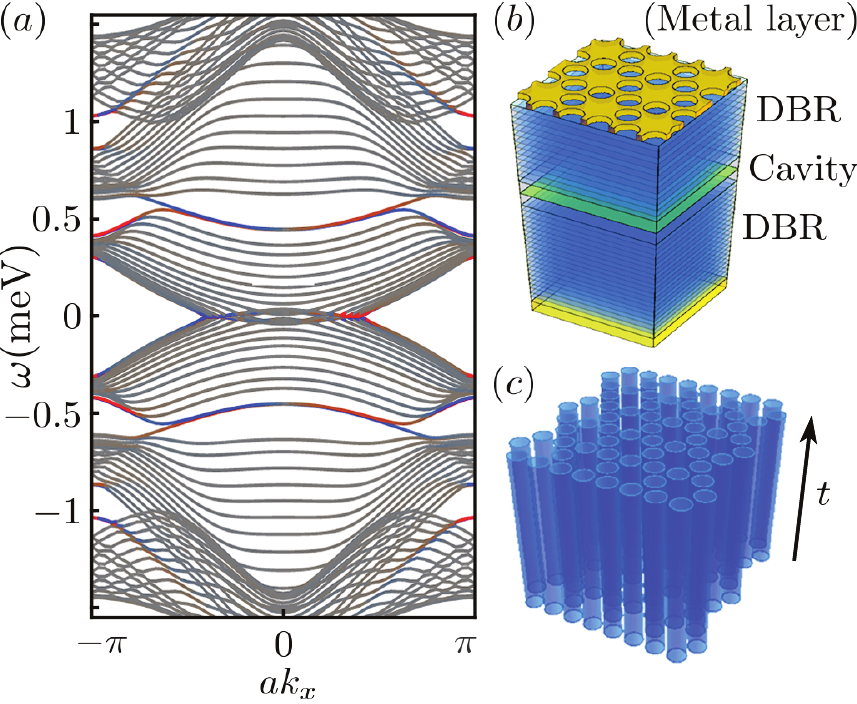}
        \caption{(a) Bogoliubov spectrum of exciton-polariton systems (in strip geometry), exhibiting similar features as the tight-binding spectrum of Fig.~\ref{fig:tightBinding}a, except for the upper topological gap which vanishes due to the modified dispersion. Although particle- and hole-like excitations overlap in energy (around $\omega = 0$), all eigenvalues are real, indicating that the system is stable. Parameters were taken from Ref.~\cite{Manni2012}, giving a low-energy topological gap of about $0.06$meV~\cite{supplementalMaterial}. (b) Practical realization in semiconductor microcavities composed of distributed Bragg reflectors (DBRs)~\cite{Masumoto2012}, with periodic potential $V(\mathbf{x})$ induced, e.g., via metal surface patterning (other mechanisms are discussed in the text). (c) Realization of the effective potential $V(\mathbf{x})$ in a nonlinear optical system using coupled waveguides~\cite{Minardi2010}.}
        \label{fig:practicalRealization}
    \end{center}
\end{figure}

The fact that polariton-polariton interactions are repulsive implies that a (Kagome) periodic potential $V(\mathbf{x})$ is required for our scheme (as discussed below Eq.~\eqref{eq:NLSE}). Such potential can be realized, e.g., by depositing thin metal films on the structure surface~\cite{Masumoto2012,Kim2013} (as illustrated in Fig.~\ref{fig:practicalRealization}b), by applying combinations of surface acoustic waves~\cite{CerdaMendez2013} or by etching micropillar arrays~\cite{Jacqmin2014}. Alternatively, one can engineer the desired potential optically~\cite{Amo2010} by taking advantage of the fact that polaritons have two possible circular polarizations (spin projections along the structure growth axis), and that polaritons with opposite spins typically interact attractively~\cite{Vladimirova2010,Takemura2014}. Specifically, one can consider Bogoliubov excitations with spin $\sigma = \pm 1$ and use a component with spin $-\sigma$ to induce the desired potential. In practice, this can be achieved by pumping both components simultaneously with an elliptically polarized incident field.

Here we follow this all-optical approach and consider Bogoliubov excitations with spin $\sigma$ on top of a polariton condensate with spin components $\phi_{0,\sigma}(\mathbf{x})$ and $\phi_{0,-\sigma}(\mathbf{x})$ (defined as in Eq.~\eqref{eq:fluctuations}). The relevant Bogoliubov equation is then given by Eq.~\eqref{eq:bogoliubovEquation} with $\alpha \equiv \alpha_1$ and $V(\mathbf{x}) = \alpha_2 |\phi_{0,-\sigma}(\mathbf{x})|^2$, where $\alpha_1$ and $\alpha_2$ denote the strength of interactions between polaritons with parallel and opposite spins, respectively~\cite{supplementalMaterial}. To provide a reliable estimate of the size of the topological gaps achievable in practice, we compute the stationary fields $\phi_{0,\pm \sigma}(\mathbf{x})$ and the spectrum of Bogoliubov excitations with spin $\sigma$ in an exact and self-consistent way, without relying on any tight-binding approximation (see Supplemental Material~\cite{supplementalMaterial}). Our results are illustrated in Fig.~\ref{fig:practicalRealization}a.

In accordance with the above tight-binding analysis, the particle-like (positive) part of the low-energy Bogoliubov spectrum exhibits three Kagome-like bands with a clear topological gap between the lowest two bands~\footnote{No higher-energy gap appears here because the third band --- flat in a tight-binding approximation --- become dispersive when longer-range hopping contributions are taken into account.}. With conservative parameters from existing experiments (taken from Ref.~\cite{Manni2012}), the topological gap reaches about $0.06$meV, which exceeds typical polariton linewidths of the order of $\mu$eV~\cite{Steger2014}. Indeed, an important feature of exciton-polaritons is that they also exhibit dissipation, as light can escape through imperfect mirrors. Dissipation can be treated by introducing a decay term into the (particle/hole) dispersion, $\pm \omega_d(-i \boldsymbol{\nabla}) \rightarrow \pm \omega_d(-i \boldsymbol{\nabla}) - i/(2\tau)$, where $\tau$ is the polariton lifetime~\cite{Carusotto2004}.

%%%%%%%%%FIGURE%%%%%%%%%
%%%%%%%%%%%%%%%%%%%%%%

%========================================================================================
% Discussion
%========================================================================================

{\it Discussion.---}A variety of other bosonic systems would be suitable for the realization of our proposal. The nonlinear Schr\"odinger equation~\eqref{eq:NLSE} provides, in particular, a direct representation of Maxwell's equations for light propagating through a nonlinear medium with polarization transverse to the propagation direction $z$ (where $t \rightarrow z$ in Eq.~\eqref{eq:NLSE} and $\mathbf{x} = (x,y)$ defines the lateral coordinates)~\cite{Kelley1965}. In that case the potential $V(\mathbf{x})$ can be induced optically. In particular, in nonlinear media with a strong electro-optic anisotropy, polarizations can be chosen such that a suitably polarized incident field experiences negligible self-nonlinearity and induces a stable potential $V(\mathbf{x})$ for optical fields with opposite (orthogonal) polarization, which experience a strong nonlinearity~\cite{Fleischer2003,Fleischer2005,Lederer2008}. Alternatively, $V(\mathbf{x})$ can be realized by refractive-index modulation, e.g., in arrays of coupled waveguides~\cite{Szameit2005,Szameit2010} as depicted in Fig.~\ref{fig:practicalRealization}c. In this context, nonlinearities $|\alpha \phi_0|^2 > 0.2 \mathrm{cm}^{-1}$ with significantly weaker losses $< 0.04 \mathrm{cm}^{-1}$ have been demonstrated using fused silica waveguides~\footnote{In Ref.~\cite{Szameit2006}, the Kerr coefficient was $n_2 = 1.35 \cdot 10^{-20} \mathrm{m}^2 \mathrm{W}^{-1}$ for a system with cross-sectional area $A \approx 160 \times 160 \mu\mathrm{m}^2$ pumped with a pulsed laser with wavelength $\lambda_0 = 800 \mathrm{nm}$. For intense but reasonable peak powers $P > 3 \cdot 10^3 \mathrm{kW}$, nonlinearities $|\alpha \phi_0|^2 = 2\pi n_0/\lambda_0 n_2 P^2 > 0.19 \mathrm{cm}^{-1}$ are obtained which are significantly larger than typical losses $< 0.04 \mathrm{cm}^{-1}$.}.

Equation~\eqref{eq:NLSE} is also relevant for the description of cold atom condensates,  providing an alternative to existing schemes for topological states. Optical flux lattices (in which optically dressed atoms both experience fluxes and a periodic potential) have been proposed and realized in this context~\cite{Cooper2011,Aidelsburger2011}, with a variety of other available methods for optical phase imprinting~\cite{Dalibard2011,Aidelsburger2013,Miyake2013,Jotzu2014}. The stirring of a cold atom condensate by rotating laser beams is also well known for realizing Abrikosov-type lattices~\cite{AboShaeer2001}. Although this would also be suitable for generating topological Bogoliubov modes (via Eq.~\eqref{eq:bogoliubovEquation}), the transfer of total angular momentum by stirring would imprint a net global flux, unlike in our scheme. In that case the Bogoliubov modes could be considered analogous to a recent proposal for topological phononic crystals, where sound waves propagate on top of a lattice of flowing air~\cite{Yang2014}.

%========================================================================================
% Conclusion
%========================================================================================

{\it Conclusion.---}We proposed a generic scheme to create topological Bogoliubov excitations in systems described by a 2D nonlinear Schr\"odinger equation --- a universal equation governing systems ranging from exciton-polariton condensates to cold atom condensates and nonlinear optical media. The key to our proposal is a background mean field exhibiting vortex-antivortex pairs and a Kagome intensity pattern, which can typically be generated by optical means. By virtue of weak interactions, Bogoliubov excitations propagating on top of this condensate acquire non-trivial phases which break TR symmetry and grant access to topological states. The resulting topological Bogoliubons manifest as robust chiral edge states, as we have explicitly demonstrated using both tight-binding and full wave-expansion methods. Our scheme does not require any external magnetic field or net transfer of orbital angular momentum, which allows to generate chiral edge modes while keeping spin/polarization degrees of freedoms degenerate and simultaneously accessible.

%========================================================================================
% Acknowledgements
%========================================================================================

T. L. thanks Y. Chong and B. Zhang for encouraging discussions. This work was funded by the Institute for Quantum Information and Matter, an NSF Physics Frontiers Center with support of the Gordon and Betty Moore Foundation (grant GBMF1250). Financial support from the Swiss National Science Foundation (SNSF) is also gratefully acknowledged.

%%%%%%%%%%%%%%%%%%%%%%%%%%%%%%%%%%%%%%%%%%%%%%%%%%%%%%%%%%%%

\bibliographystyle{apsrev4-1}
\bibliography{manuscript}

%%%%%%%%%%%%%%%%%%%%%%%%%%%%%%%%%%%%%%%%%%%%%%%%%%%%%%%%%%%%
\end{document}